\documentclass[pra,aps,floatfix,twocolumn,superscriptaddress,amsmath,amssymb]{revtex4-2}
\usepackage{graphicx}
\usepackage{xcolor}
\usepackage{bm}
\usepackage{hyperref}
\hypersetup{colorlinks=true,citecolor=blue,linkcolor=blue,urlcolor=blue}
\usepackage[T1]{fontenc}
\usepackage{newtxtext,newtxmath}
\usepackage[normalem]{ulem}
\allowdisplaybreaks

\begin{document}

\title{%
Entanglement entropy dynamics of non-Gaussian states
in free boson systems: \\
Random sampling approach
}

\author{Ryui Kaneko}
\email{ryuikaneko@sophia.ac.jp}
\affiliation{%
Physics Division, Sophia University, Chiyoda, Tokyo 102-8554, Japan}

\author{Daichi Kagamihara}
\email{dkagamihara119@g.chuo-u.ac.jp}
\affiliation{%
Department of Physics, Chuo University, Bunkyo, Tokyo 112-8551, Japan}

\author{Ippei Danshita}
\email{danshita@phys.kindai.ac.jp}
\affiliation{%
Department of Physics, Kindai University, Higashi-Osaka, Osaka 577-8502, Japan}

\date{\today}

\begin{abstract}
We develop a random sampling method for
calculating the time evolution of the R\'{e}nyi entanglement entropy
after a quantum quench from an insulating state in free boson systems.
Because of the non-Gaussian nature of the initial state,
calculating the R\'{e}nyi entanglement entropy
calls for the exponential cost of computing a matrix permanent.
We numerically demonstrate that
a simple random sampling method reduces
the computational cost of a permanent;
for an $N_{\mathrm{s}}\times N_{\mathrm{s}}$ matrix
corresponding to $N_{\mathrm{s}}$ sites at half filling,
the sampling cost becomes $\mathcal{O}(2^{\alpha N_{\mathrm{s}}})$
with a constant $\alpha\ll 1$,
in contrast to the conventional algorithm
with the $\mathcal{O}(2^{N_{\mathrm{s}}})$
number of summations requiring
the exponential time cost.
Although the computational cost is still exponential,
this improvement allows us to obtain the entanglement entropy dynamics
in free boson systems for more than $100$ sites.
We present several examples of the entanglement entropy dynamics in
low-dimensional free boson systems.
\end{abstract}

\maketitle
 
\section{Introduction}
\label{sec:intro}

Understanding the dynamics of quantum many-body systems
is a central issue in modern physics.
The entanglement entropy is a key quantity
that characterizes
the dynamics of quantum many-body systems.
For example, temporal entanglement entropy can signal quantum phase
transitions, helping us identify and understand new phases of matter in
nonequilibrium quantum systems~\cite{heyl2018}.
Additionally, it provides insights into
how information flows and evolves in quantum
systems~\cite{zyczkowski2001}.
Since entanglement
is a crucial resource for quantum computing and cryptography,
understanding its dynamics may lead to the development of efficient
quantum algorithms and secure communication
protocols~\cite{zyczkowski2001,nielsen2010}.
These studies motivate us to investigate
the thermalization process
in quantum many-body systems
and the propagation of quantum
information~\cite{greiner2002,
calabrese2005,
dechiara2006,
cramer2008a,
cramer2008b,
flesch2008,
fagotti2008,
lauchli2008,
eisert2010,
barmettler2012,
bardarson2012,
cheneau2012,
trotzky2012,
carleo2014,
bauer2015,
frerot2015,
frerot2016,
laflorencie2016,
alba2017,
alba2018,
goto2019,
nagao2019,
yao2020,
takasu2020,
kunimi2021,
nagao2021,
yoshii2022,
rylands2022,
kaneko2022,
kagamihara2023,
yamashika2023,
yamashika2024_arxiv}.
Although the von Neumann entanglement entropy is not a directly measurable quantity,
there are several proposals to measure
R\'enyi entanglement entropy~\cite{abanin2012,
daley2012,elben2018}.
Recent experiments have successfully observed
the dynamics of the R\'enyi entanglement entropy
using ultracold atoms in optical
lattices~\cite{islam2015,
kaufman2016} and trapped ions~\cite{brydges2019}.

The numerical simulation of dynamics of the entanglement entropy
is also an important approach to understanding quantum many-body
systems and providing a benchmark for experiments.
Studying the efficiency of numerical simulations provides insights
into problems that are challenging in classical systems but can be
effectively addressed in quantum systems~\cite{watrous2009}.
In some equilibrium systems,
the R\'enyi entanglement entropy can be efficiently calculated using
quantum Monte Carlo simulations~\cite{hastings2010,zhao2022}.
This approach allows for accurate
entanglement measurements in large and complex systems, shedding light
on their universal properties. However, this efficiency in classical
simulations does not always extend to nonequilibrium and general
equilibrium quantum systems.
Finding efficient simulation methods for such systems
would guide which problems
are most suitable for applying
digital and analog quantum simulations.

In contrast to fermion and spin systems,
boson systems are much harder to simulate
because of the large number of local Hilbert spaces.
Even in the free boson systems with simple initial states,
such as the Mott insulating state and the charge-density-wave (CDW) state,
the entanglement entropy dynamics is difficult to calculate
because of the non-Gaussian nature of the initial states.
Although the analytical formula for the entanglement entropy
is formally obtained by a matrix permanent,
its numerical evaluation requires the exponential cost~\cite{kagamihara2023}.
This situation limits the system size that can be studied
to a few tens of sites or particles.
Therefore, understanding dynamics of the entanglement entropy
in boson systems remains to be a challenging problem
even in noninteracting systems.

In this paper, we develop a random sampling method
for calculating the time evolution of the R\'{e}nyi entanglement entropy
in free boson systems.
In the developed method, we still need to evaluate the matrix permanent,
which requires the exponential computational cost in general.
However, the growth rate
of the computational cost is much slower than the exact permanent calculation.
We numerically found that
the computational cost is reduced to $\mathcal{O}(2^{\alpha N_{\mathrm{s}}})$
with a small constant $\alpha\ll 1$ and the system size $N_{\mathrm{s}}$.
To be more specific,
we calculated the size-dependent statistical error
of the entanglement entropy,
which scales as $\sqrt{c/N_{\mathrm{total}}}$,
with $c$ being a size-dependent constant
and $N_{\mathrm{total}}$ being the total number of samples,
and investigated how $c$ grows with the system size,
which represents the number of samples required to achieve a given
statistical error.
The constant $\alpha$ for the sampling cost is defined by
\begin{align}
 c = 2^{\alpha N_{\mathrm{s}} + \mathrm{const.}}.
\end{align}
This improvement enables us to study dynamics of the entanglement entropy
in free boson systems for more than $100$ sites,
confirming that the entanglement entropy in the long-time region
exhibits the volume-law scaling as expected.

This paper is organized as follows.
In Sec.~\ref{sec:method},
we briefly review the calculation of the
R\'{e}nyi entanglement entropy in free boson systems
and describe the conventional algorithm for evaluating the entanglement
entropy, which requires the computation of a matrix permanent.
To reduce the computational cost, we propose a random sampling
method for the matrix permanent.
In Sec.~\ref{sec:results},
we examine the performance of the random sampling method
by estimating the size dependence of the statistical error.
We then present numerical results for dynamics of the entanglement entropy
in free boson systems for spatial one (1D) and two dimensions (2D).
Finally, in Sec.~\ref{sec:conclusions},
we summarize our results and discuss future prospects.
For simplicity,
we set $\hbar = 1$ and
take the lattice constant to be unity
throughout this paper.

\section{Random sampling method for entanglement entropy}
\label{sec:method}

We first briefly review how to evaluate time evolution of the R\'{e}nyi entanglement
entropy in free boson
systems in the case that the initial state is an insulating state.
Let us consider dynamics subjected to a quantum quench in the Bose-Hubbard model
under the open boundary condition.
The Hamiltonian is defined as
\begin{align}
\label{eq:ham}
 \hat{H}
 = - J \sum_{\langle j,l\rangle} ( \hat{b}^{\dagger}_j \hat{b}_{l} + \mathrm{H.c.} )
 + \sum_{j} \Omega_j \hat{n}_j
 + \frac{U}{2} \sum_{j} \hat{n}_j (\hat{n}_j-1),
\end{align}
where the symbols $\hat{b}_j$ and $\hat{n}_j$ are
the boson annihilation and number operators, respectively.
The parameters $J$, $U$, and $\Omega_j$ represent
the strength of the hopping,
the strength of the interaction,
and
the single-particle potential, respectively.
The symbol $\langle j,l\rangle$ means that sites $j$ and $l$ are nearest neighbors.

We focus on a sudden quench from an insulating state
to the noninteracting ($U=0$) and homogeneous ($\Omega_j = 0$) point.
The following discussion may also be applicable to the case of a quench
to the noninteracting and inhomogeneous point;
however, we do not consider such a case in this paper for simplicity.
The quench to the interacting point is also interesting,
but the system becomes nonintegrable
and it is beyond the scope of this paper.

As an initial state,
we specifically choose
the $010101\cdots$-type CDW state at half filling.
Although the following formalism also holds for any Fock initial state
with any noninteracting Hamiltonian after the quench,
we focus on the CDW state for simplicity.
It is defined as
\begin{align}
\label{eq:cdw_init}
 |\psi\rangle
 = \prod_{j \in \mathrm{G_{CDW}}} \hat{b}^{\dagger}_{j} |0\rangle,
\end{align}
where $|0\rangle$ is the vacuum state of $\hat{b}_j$.
The $\mathrm{G_{CDW}}$ corresponds to the set of charge rich sites.
For example,
$\mathrm{G_{CDW}} = \{2,4,6,\dots\}$ in 1D,
and
$\mathrm{G_{CDW}} = \{(2,1),(4,1),(6,1),\dots$
$(1,2),(3,2),(5,2), \dots$
$(2,3),(4,3),(6,3),\dots$
$(1,4),(3,4),(5,4), \dots\}$
in 2D,
respectively.
Hereafter, in 2D, we map the site index $j (= j_x + L_x j_y)$
one-to-one to the lattice site $(j_x,j_y)$
for $j_x=1,2,\dots,L_x$ and $j_y=1,2,\dots,L_y$
on a square lattice
with $L_x$ ($L_y$) being the length of the side
along the $x$ ($y$) direction, respectively.
The number of sites is represented as $N_{\mathrm{s}}$,
which is taken as an even number in our study.
Then, the number of particles is $N_{\mathrm{b}}=N_{\mathrm{s}}/2$.
The CDW state can be obtained as the ground state of the Bose-Hubbard model
at half filling
for $\Omega/J \gg 1$ and $U / J \gg 1$
when $\Omega_j = \Omega(-1)^{j+1}$ in 1D
and $\Omega_j = \Omega(-1)^{j_x+j_y}$ in 2D,
respectively.
One can prepare the CDW state in experiments using a secondary optical lattice,
which has a lattice constant twice as large as that of the primary
lattice~\cite{trotzky2012}.
Note that such CDW states and also the Mott insulating state
that appear in the Bose-Hubbard model are non-Gaussian states,
although the counterparts in the Fermi-Hubbard model
are Gaussian states.

To make the discussion self-contained,
we summarize the calculation of the R\'{e}nyi entanglement entropy
in free boson systems~\cite{kagamihara2023}.
We previously evaluated
the second
R\'{e}nyi entanglement entropy,
which is defined by
\begin{align}
 S_2(t) = -\ln \mathrm{Tr}_{\mathrm{G}}
 \left[\hat{\rho}_{\mathrm{G}}(t)\right]^2,
\end{align}
for 
the time-evolved state,
$|\psi(t)\rangle = \exp(-i\hat{H}t) |\psi\rangle$~\cite{kagamihara2023}.
Here, $\hat{\rho}_{\mathrm{G}}(t)$ is the reduced density matrix
and $\mathrm{Tr}_{\mathrm{G}}$ is the trace over the basis of
subsystem $\mathrm{G}$
that contains
$l=1$, $2$, $\dots$, $N_{\mathrm{G}}$ sites.
For simplicity, we set $N_{\mathrm{G}}$ to half the system size
($N_{\mathrm{G}} = N_{\mathrm{s}}/2$) throughout this paper.
The reduced density matrix
$\hat{\rho}_{\mathrm{G}}(t)$
and 
the product of two copies of the state $|\psi(t)\rangle$,
i.e.,
$|\psi_{\mathrm{copy}}(t)\rangle
:= |\psi(t)\rangle \otimes |\psi(t)\rangle$,
are related as
\begin{align}
 \mathrm{Tr}_{\mathrm{G}}
 \left[\hat{\rho}_{\mathrm{G}}(t)\right]^2
 =
 \langle \psi_{\mathrm{copy}}(t) |
 \hat{V}_{\mathrm{G}}
 | \psi_{\mathrm{copy}}(t) \rangle,
\end{align}
with $\hat{V}_{\mathrm{G}}(t)$ being the shift operator
that swaps states in subsystem $\mathrm{G}$.
Therefore, we need to evaluate the right-hand side of the above equation
by explicitly calculating the time-evolved state
$|\psi(t)\rangle$.
For any noninteracting Hamiltonian $\hat{H}_0$,
by diagonalizing it in the first-quantization representation,
we can express it as
\begin{align}
\label{eq:ham_0}
 \hat{H}_0
 =
 - J \sum_{\langle j,l\rangle} ( \hat{b}^{\dagger}_j \hat{b}_{l} + \mathrm{H.c.} )
 =
 \sum_{k=1}^{N_{\mathrm{s}}}
 \epsilon_k \hat{\beta}^{\dagger}_k \hat{\beta}_k,
\quad
 \hat{\beta}_k
 =
 \sum_{j=1}^{N_{\mathrm{s}}}
 x_{k,j} \hat{b}_j,
\end{align}
where $\hat{b}_j$ is the annihilation operator
in the original basis of the Hamiltonian
and $\hat{\beta}_k$ is the annihilation operator
in the basis diagonalizing the Hamiltonian.
The $k$th eigenenergy of $\hat{H}_0$ is represented as $\epsilon_k$
and the corresponding eigenvector is expressed as $\bm{x}_k$.
The elements of the eigenvector $\bm{x}_k$ are real numbers
when the hopping strength $J$ is real.
Straightforward calculations for $\hat{H}=\hat{H}_0$
lead to the expression of the
time-evolved state as
\begin{align}
 |\psi(t)\rangle
 &=
 \exp(-i\hat{H}_0 t)
 \prod_{j \in \mathrm{G_{CDW}}}
 \hat{b}^{\dagger}_j |0\rangle
\\
 &=
 \prod_{j \in \mathrm{G_{CDW}}}
 \left[
 \exp(-i\hat{H}_0 t) \hat{b}^{\dagger}_j
 \exp(i\hat{H}_0 t) 
 \right]
 |0\rangle
\\
 &=
 \prod_{j \in \mathrm{G_{CDW}}}
 \left\{
 \sum_{j'=1}^{N_{\mathrm{s}}}
 \left[
 \sum_{k=1}^{N_{\mathrm{s}}}
 x_{k,j} x_{k,j'} \exp(-i\epsilon_k t)
 \right]
 \hat{b}^{\dagger}_{j'}
 \right\}
 |0\rangle.
\end{align}
Here, we use the fact that
$\exp(\pm i\hat{H}_0 t) |0\rangle = |0\rangle$.
For convenience, we define the correlation $y_{i,j}(t)$
($i$, $j=1$, $2$, $\dots$, $N_{\mathrm{s}}$)
as
\begin{align}
\label{eq:ee_y}
 y_{i,j}(t)
 =
 \sum_{k=1}^{N_{\mathrm{s}}}
 x_{k,i} x_{k,j} \exp(-i\epsilon_k t).
\end{align}
Then, the time-evolved state is expressed as
\begin{align}
 |\psi(t)\rangle
 =
 \prod_{j \in \mathrm{G_{CDW}}}
 \left\{
 \sum_{j'=1}^{N_{\mathrm{s}}}
 y_{j,j'}(t) \hat{b}^{\dagger}_{j'}
 \right\}
 |0\rangle.
\end{align}
The state $|\psi_{\mathrm{copy}}(t)\rangle$ is obtained as
a tensor product of two $|\psi(t)\rangle$ states.
Because both $|\psi_{\mathrm{copy}}(t)\rangle$
and $\hat{V}_{\mathrm{G}} |\psi_{\mathrm{copy}}(t)\rangle$
are many-boson states
and their wave functions are symmetric under the exchange of bosons,
the expectation value of the shift operator
$
 \langle \psi_{\mathrm{copy}}(t) |
 \hat{V}_{\mathrm{G}}
 | \psi_{\mathrm{copy}}(t) \rangle
$ 
is given by the permanent of a certain matrix
consisting of
single-particle correlation functions,
defined by
\begin{align}
\label{eq:ee_z}
 z_{i,j}(t)
 =
 \sum_{l\in \mathrm{G}} y^{*}_{r_i,l}(t) y_{r_j,l}(t),
\end{align}
with $r_i$ and $r_j$ being indices of charge rich sites
($r_i, r_j \in \mathrm{G_{CDW}}$).
Consequently, using the matrix $Z$ with elements $z_{i,j}(t)$,
we can express the R\'{e}nyi entanglement entropy at time $t$ as
\begin{align}
\label{eq:ee_sq}
 S_2 &= -\ln \mathrm{perm} A,
\\
\label{eq:ee_a}
 A &=
 \begin{pmatrix}
 Z & I-Z \\
 I-Z & Z \\
 \end{pmatrix}.
\end{align}
Here,
$\mathrm{perm} A$ is the matrix permanent,
which is defined as the sum of all the products of the elements of the
matrix $A$, given as
\begin{align}
\label{eq:def_perm}
 \mathrm{perm} A
 =
 \sum_{f \in \frak{S}_{N_{\mathrm{s}}}}
 \prod_{j=1}^{N_{\mathrm{s}}} a_{j,f(j)},
\end{align}
with $\frak{S}_{N_{\mathrm{s}}}$ being the set of all permutations.
Then,
$A$ is an $N_{\mathrm{s}}\times N_{\mathrm{s}}$ square matrix,
$I$ is an $N_{\mathrm{s}}/2\times N_{\mathrm{s}}/2$ identity matrix, and
$Z$ is an $N_{\mathrm{s}}/2\times N_{\mathrm{s}}/2$ matrix with elements $z_{i,j}(t)$.

Note that the matrix $Z$
takes the form in Eq.~\eqref{eq:ee_z}
for any initial Fock state and any quadratic
Hamiltonian~\cite{kagamihara2023}.
In this paper, we focus on the CDW initial state
and the free boson Hamiltonian
with the nearest-neighbor hopping
on a chain and a square lattice.
In 1D, the matrix representation of $\hat{H}_0$
in the basis of $\hat{b}_j$ is given by
\begin{align}
 h_{0,j,l}
 =
\begin{cases}
 -J & (|j-l|=1),
\\
 0 & (\text{otherwise}),
\end{cases}
\end{align}
which is an $N_{\mathrm{s}}\times N_{\mathrm{s}}$ tridiagonal matrix.
The eigenvalues and eigenvectors are easily obtained as
\begin{align}
 \epsilon_k
 &=
 - 2J \cos \left( \frac{k\pi}{N_{\mathrm{s}}+1} \right),
\\
 x_{k,l}
 &=
 \sqrt{\frac{2}{N_{\mathrm{s}}+1}}
 \sin \left( \frac{k\pi}{N_{\mathrm{s}}+1} l \right),
\end{align}
where $k, l = 1, 2, \dots, N_{\mathrm{s}}$.
In 2D, the matrix representation of $\hat{H}_0$
in the basis of $\hat{b}_j$ is given by
\begin{align}
 h_{0,j,l}
 =
\begin{cases}
 -J & (\sqrt{(j_x-l_x)^2 + (j_y-l_y)^2} = 1),
\\
 0 & (\text{otherwise}),
\end{cases}
\end{align}
where $j=j_x+L_x j_y$ and $l=l_x+L_x l_y$.
This $N_{\mathrm{s}}\times N_{\mathrm{s}}$ matrix is no longer
tridiagonal.
We numerically diagonalize the matrix $h_0$ to obtain
the eigenvalues $\epsilon_k$ and eigenvectors $\bm{x}_k$.

In general, the calculation of the matrix permanent
in Eq.~\eqref{eq:ee_sq} requires
the exponential cost of $\mathcal{O}(N_{\mathrm{s}}\times 2^{N_{\mathrm{s}}})$
for an $N_{\mathrm{s}}\times N_{\mathrm{s}}$ matrix.
The well-known algorithms for evaluating matrix permanents are
the Ryser formula~\cite{ryser1963,brualdi1991,glynn2010}
and Balasubramanian-Bax-Franklin-Glynn (BBFG)
formula~\cite{balasubramanian1980,bax1996,bax1998,glynn2010,glynn2013}.
For example, in the BBFG algorithm,
the permanent for an $n\times n$ matrix ($A$) is evaluated as
\begin{align}
\label{eq:bbfg}
 \mathrm{perm} A
 =
 \frac{1}{2^{n-1}}
 \sum_{\bm{\delta}}
 \left(\prod_{k=1}^{n} \delta_k\right)
 \prod_{j=1}^{n} \sum_{i=1}^{n} \delta_i a_{ij}.
\end{align}
Here, $a_{ij}$ is the element of the matrix $A$,
and the summation is taken over
$\bm{\delta}=(\delta_1,\delta_2,\dots,\delta_n) \in \{\pm 1\}^n$
with $\delta_1=1$.
The exponential time cost stems from the fact that
the number of terms in the summation grows exponentially in $n$.

To evaluate the permanent of an $n\times n$ matrix $A$ more efficiently,
we propose a random sampling method.
Instead of taking all the terms in the summation in Eq.~\eqref{eq:bbfg},
we randomly sample a subset of terms.
We replace the sum over the vector $\bm{\delta}$ in Eq.~\eqref{eq:bbfg}
by the random vector $\bm{r}$.
Note that the equivalent sampling procedure itself is proposed
in Ref.~[\onlinecite{huh2025}],
although the context is different and the efficiency
of the random sampling that we adopt here is not discussed.
As we show below,
this method allows us to approximately evaluate the permanent
with the matrix size larger than $100$,
which is difficult to achieve with the conventional
Ryser and BBFG algorithms.

To simplify the notation,
let us introduce the Glynn estimator
for an $n\times n$ complex matrix $A$
and the complex vector $\bm{x}$~\cite{gurvits2005,glynn2010,aaronson2014,berkowitz2018},
which is defined as
\begin{align}
\label{eq:gly}
 \mathrm{Gly}_{\bm{x}}(A)
 =
 \prod_{k=1}^{n} x^{*}_k \prod_{i=1}^{n}
 \left( \sum_{j=1}^{n} a_{ij} x_j \right).
\end{align}
When we specifically choose a random variable $\bm{r}$, which has elements
$r_i \in \mathbb{C}$ ($i=1$, $2$, $\dots$, $n$)
that are independently chosen uniformly
on $|r_i|=1$, we can evaluate the permanent of the matrix as
the expectation value of the Glynn estimator~$\mathrm{Gly}_{\bm{r}}(A)
$~\cite{gurvits2005,glynn2010,aaronson2014,berkowitz2018}.
The relation is given as
\begin{align}
 \mathrm{perm} A
 = \mathbb{E}[\mathrm{Gly}_{\bm{r}}(A)]
 = \mathbb{E}\left[\prod_{i=1}^{n} r^{*}_i
 \left( \sum_{j=1}^{n} a_{ij} r_j \right) \right],
\end{align}
where $\mathbb{E}$ means the expectation value.
This equation can be shown by expanding the product:
\begin{align}
 &~\phantom{=}~
 \mathbb{E}[\mathrm{Gly}_{\bm{r}}(A)]
\nonumber
\\
 &=
 \mathbb{E}\Biggl[
 \prod_{i=1}^{n}
 r^{*}_i
 \left(
 a_{i1} r_1 + a_{i2} r_2 + \cdots + a_{in} r_n
 \right)
 \Biggr]
\\
 &=
 \mathbb{E}\Biggl[
 ( r^{*}_1 r^{*}_2 \cdots r^{*}_n )
 \prod_{i=1}^{n}
 \left(
 a_{i1} r_1 + a_{i2} r_2 + \cdots + a_{in} r_n
 \right)
 \Biggr]
\\
 &=
 \mathbb{E}\Biggl[
 ( r^{*}_1 r^{*}_2 \cdots r^{*}_n )
 \nonumber
 \\
 &~\phantom{==}~\times
 \sum_{g(1),g(2),\dots,g(n) \in \{1,2,\dots,n\}}
 a_{1g(1)} a_{2g(2)} \cdots a_{ng(n)}
 \nonumber
 \\
 &~\phantom{===}~\times
 ( r_{g(1)} r_{g(2)} \cdots r_{g(n)} )
 \Biggr]
\\
 &=
 \sum_{g(1),g(2),\dots,g(n) \in \{1,2,\dots,n\}}
 a_{1g(1)} a_{2g(2)} \cdots a_{ng(n)}
 \nonumber
 \\
 &~\phantom{==}~\times
 \mathbb{E}\Bigl[
 ( r^{*}_1 r^{*}_2 \cdots r^{*}_n )
 ( r_{g(1)} r_{g(2)} \cdots r_{g(n)} )
 \Bigr].
\end{align}
If the map $i\mapsto g(i)$ is a permutation,
one can always find
a unique pairing between $r^{*}_i$ and $r_{g(j)}$
for all $i[=g(j)]$ and $j$, and
the expectation value satisfies
\begin{align}
 &~\phantom{=}~
 \mathbb{E}\Bigl[( r^{*}_1 r^{*}_2 \cdots r^{*}_n )
 ( r_{g(1)} r_{g(2)} \cdots r_{g(n)} )\Bigr]
\nonumber
\\
 &=
 \mathbb{E}\Bigl[ |r_1|^2 |r_2|^2 \cdots |r_n|^2 \Bigr]
 = \mathbb{E}\Bigl[ 1^n \Bigr]
 = 1;
\end{align}
otherwise it is zero
because there exists at least one unpaired and independent
$r^{*}_i$ that satisfies
$\mathbb{E}[r^{*}_i]=0$.
Then, we obtain
\begin{align}
 \mathbb{E}[\mathrm{Gly}_{\bm{r}}(A)]
 =
 \sum_{g(1),g(2),\dots,g(n) \in \frak{S}_n}
 a_{1g(1)} a_{2g(2)} \cdots a_{ng(n)},
\end{align}
which gives the permanent of the matrix $A$
in Eq.~\eqref{eq:def_perm}.
Therefore,
we can calculate the permanent
by the following sample mean:
\begin{align}
\label{eq:rnd_smp_prod_rar}
 \mathrm{perm} A
 &\approx
 \frac{1}{N_{\mathrm{smp}}}
 \sum_{m=1}^{N_{\mathrm{smp}}}
 p^{(m)},
 \\
\label{eq:rnd_smp_prod_rar_p}
 p^{(m)}
 &=
 \prod_{i=1}^{n} {r^{(m)}_i}^{*}
 \left( \sum_{j=1}^{n} a_{ij} r^{(m)}_j \right),
\end{align}
where $N_{\mathrm{smp}}$ is the number of samples,
and $\bm{r}^{(m)}$ is a complex random vector of a sample $m$.
In practice, we take $r_i^{(m)}=\exp[i\theta_i^{(m)}]$
with $\theta_i^{(m)}$ chosen uniformly in
$[0,2\pi)$~\cite{aaronson2014,berkowitz2018,kocharovsky2020}.
The value $p^{(m)}$ is a complex number for each sample $m$.

In the present system with $n=N_{\mathrm{s}}$,
the entanglement entropy satisfies
$0\le S_2\le c N_{\mathrm{s}}$ with $c$ being a sufficiently large constant,
and therefore,
the condition
$\exp(- c N_{\mathrm{s}}) \le \mathrm{perm} A \le 1$
holds.
Since $\mathrm{Re}\, p^{(m)}$ and $\mathrm{Im}\, p^{(m)}$
can be exponentially small in $N_{\mathrm{s}}$
and can be both positive and negative,
we need $N_{\mathrm{smp}}=\mathcal{O}[\exp(\alpha N_{\mathrm{s}})]$
samples with a constant $\alpha$
to accurately estimate $\mathrm{perm} A$ in general.
The situation is similar to the case of systems
having the notorious negative sign problems~\cite{loh1990}.
The advantage of the present approach is that
the constant prefactor $\alpha$ would be sufficiently smaller than unity
as far as we deal with the matrix $A$ generated in the present system
[see Eq.~\eqref{eq:ee_a}].
Indeed, we numerically found that $\alpha\approx 0.2$.
One may also consider the importance sampling
to reduce the variance of the estimator
in Eq.~\eqref{eq:gly}.
However, we do not use it in the present study
and stick to the simple random sampling method
because we are interested in
how far we can go with the primitive procedure.
Note that, although we here focus on the CDW initial state to be specific,
the approach described above is applicable also to other insulating
initial states as long as they are expressed as a simple product of
local Fock states.

\section{Results}
\label{sec:results}

Hereafter, we present the numerical results
on dynamics of the entanglement entropy
in hypercubic lattices,
such as a chain in 1D and a square lattice in 2D.
The number of sites is given by
$N_{\mathrm{s}}$ on a one-dimensional chain
and
$N_{\mathrm{s}} = L_x \times L_y$ on a two-dimensional square lattice.

\subsection{Estimation of the statistical error}
\label{subsec:estimation_error}

\begin{figure}[!t]
\centering
\includegraphics[width=\columnwidth]{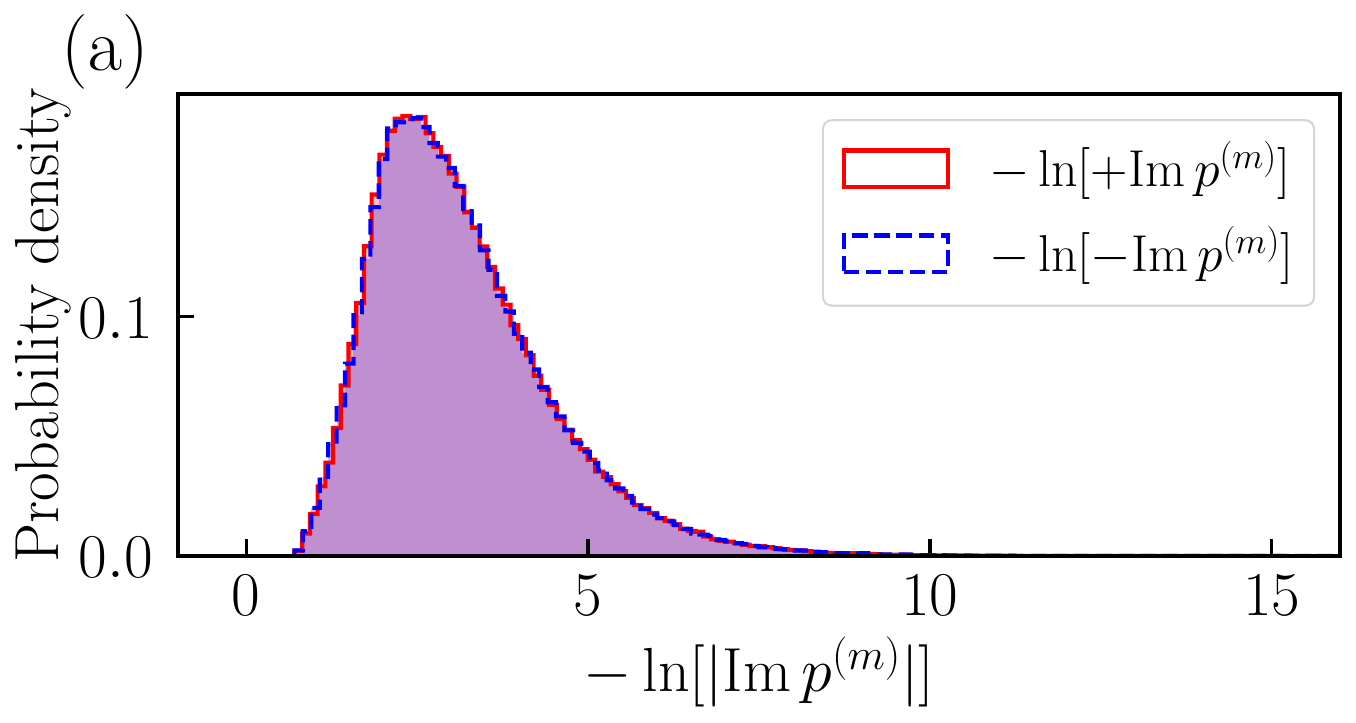}
\\
\includegraphics[width=\columnwidth]{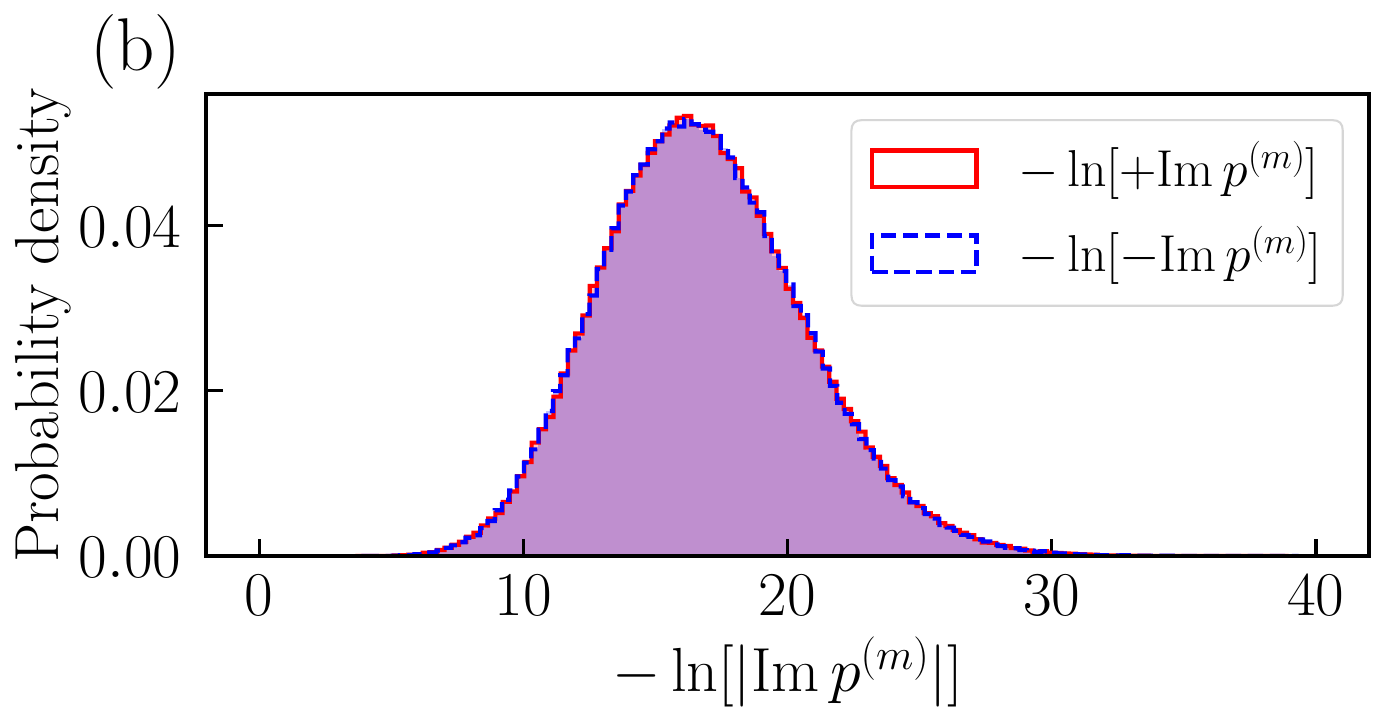}
\caption{Distribution of the value
$-\ln[ |\mathrm{Im}\, p^{(m)}| ]$
in Eq.~\eqref{eq:rnd_smp_prod_rar_p}.
We show the distribution $P(x)$ of $x=-\ln[+\mathrm{Im}\, p^{(m)}]$
($x=-\ln[-\mathrm{Im}\, p^{(m)}]$)
when $\mathrm{Im}\, p^{(m)}>0$ ($\mathrm{Im}\, p^{(m)}<0$) with a red solid line (a blue dashed line).
(a) At time $tJ=1$ for $N_{\mathrm{s}}=40$.
(b) At time $tJ=2N_{\mathrm{s}}$ for $N_{\mathrm{s}}=40$.
In both cases, the positive and negative components exhibit
nearly the same distribution,
suggesting that $\mathrm{perm} A$ does not contain an imaginary part.
}
\label{fig:1}
\end{figure}

\begin{figure}[!t]   
\centering
\includegraphics[width=\columnwidth]{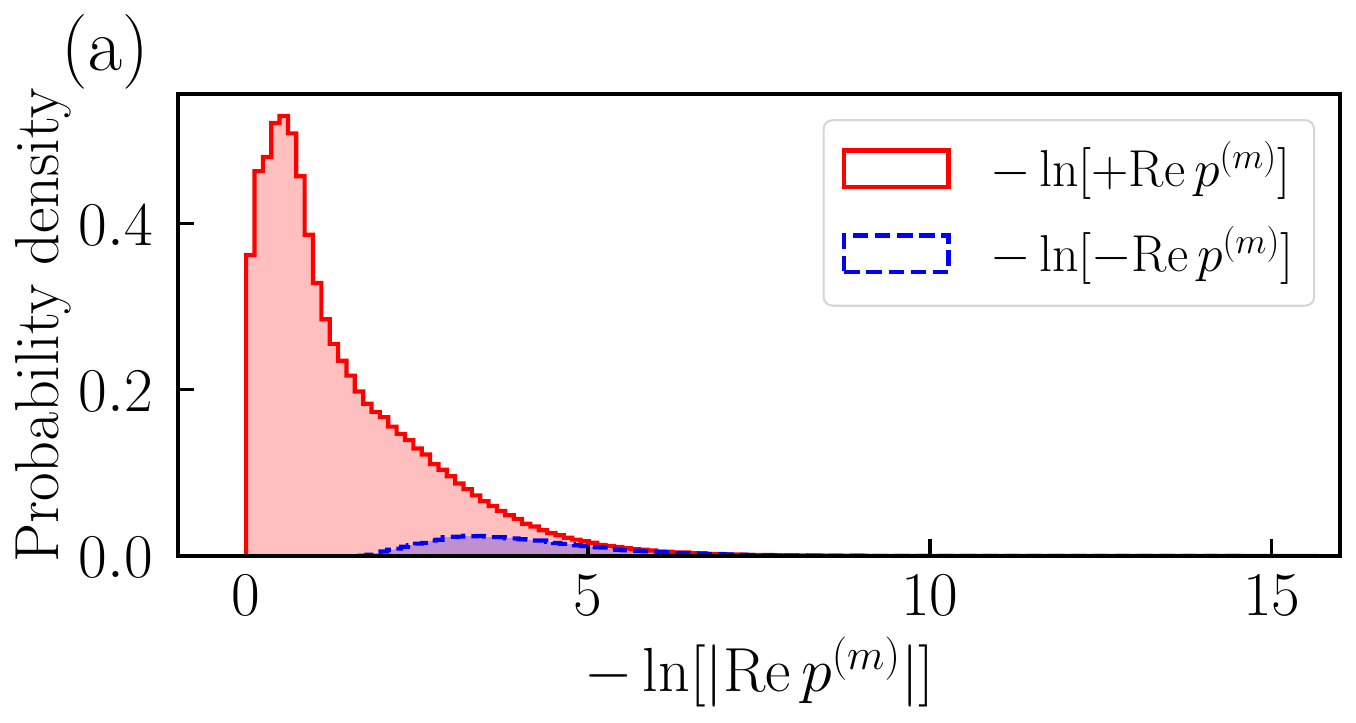}
\\
\includegraphics[width=\columnwidth]{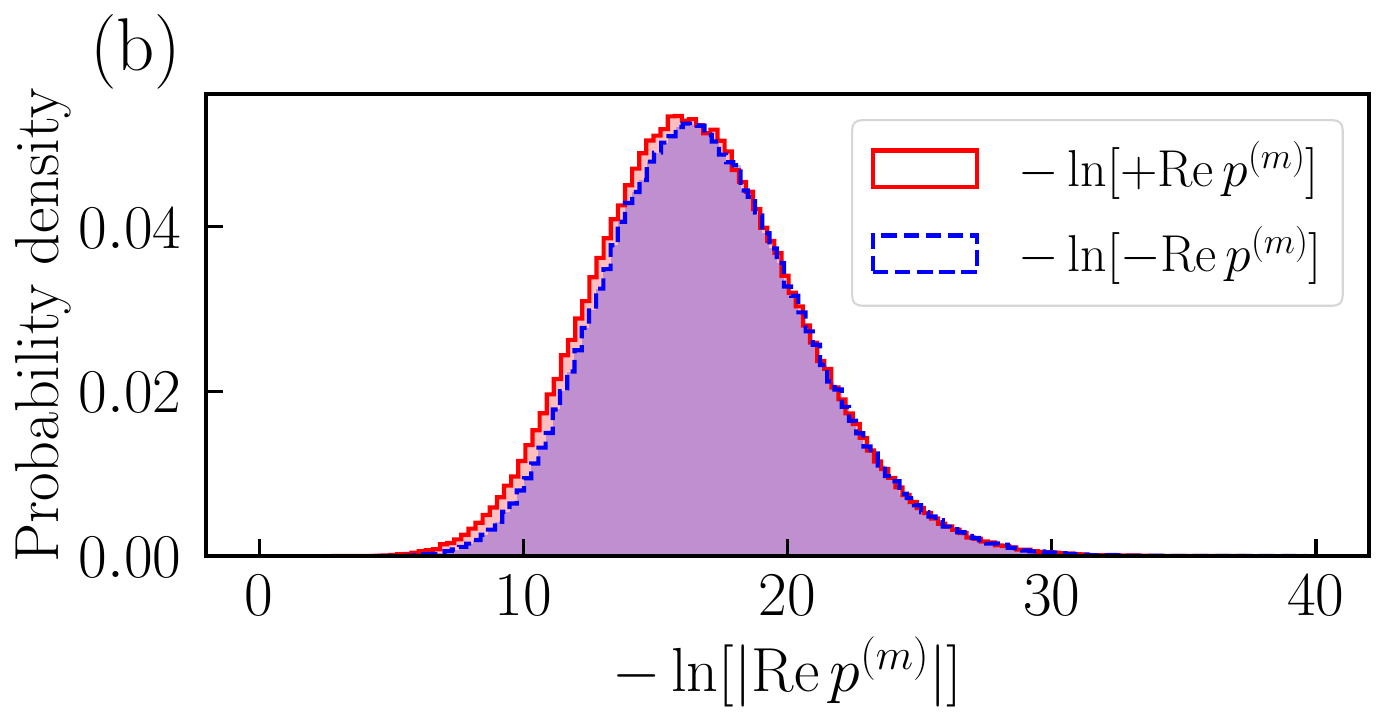}
\caption{Distribution of the value
$-\ln[ |\mathrm{Re}\, p^{(m)}| ]$
in Eq.~\eqref{eq:rnd_smp_prod_rar_p}.
We show the distribution $P(x)$ of $x=-\ln[+\mathrm{Re}\, p^{(m)}]$   
($x=-\ln[-\mathrm{Re}\, p^{(m)}]$)
when $\mathrm{Re}\, p^{(m)}>0$ ($\mathrm{Re}\, p^{(m)}<0$) with a red solid line (a blue dashed line).
(a) At time $tJ=1$ for $N_{\mathrm{s}}=40$.
Since the positive component is dominant,
we expect $\mathrm{perm} A=\mathcal{O}(1)$,
and thus, $S_2=\mathcal{O}(1)$.
(b) At time $tJ=2N_{\mathrm{s}}$ corresponding to
$S_2=\mathcal{O}(N_{\mathrm{s}})$ for $N_{\mathrm{s}}=40$.
Since the positive and negative components are comparable
while the positive one is slightly dominant,
we expect $\mathrm{perm} A=\mathcal{O}[\exp(-\mathrm{const.}\times
N_{\mathrm{s}})]$,
and thus, $S_2=\mathcal{O}(N_{\mathrm{s}})$.
}
\label{fig:2}
\end{figure}

\begin{figure}[!t]
\centering
\includegraphics[width=\columnwidth]{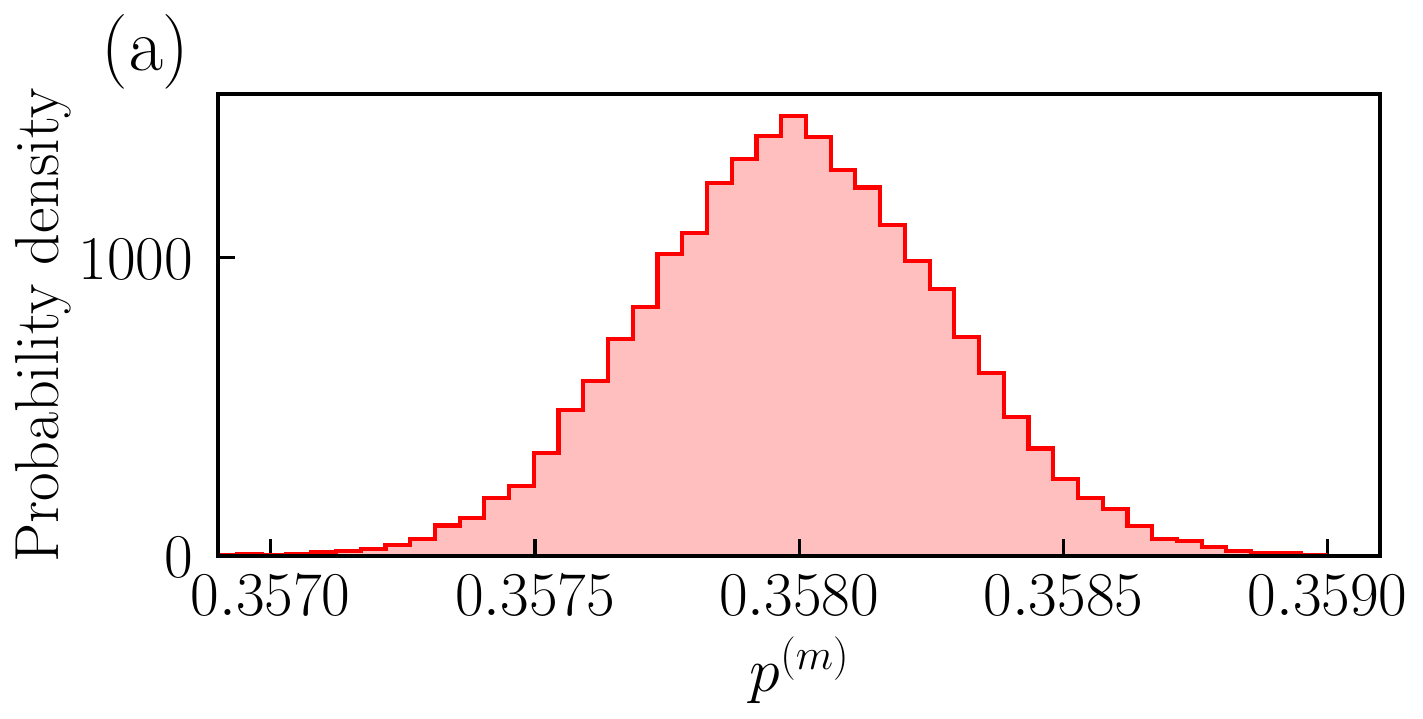}
\\
\includegraphics[width=\columnwidth]{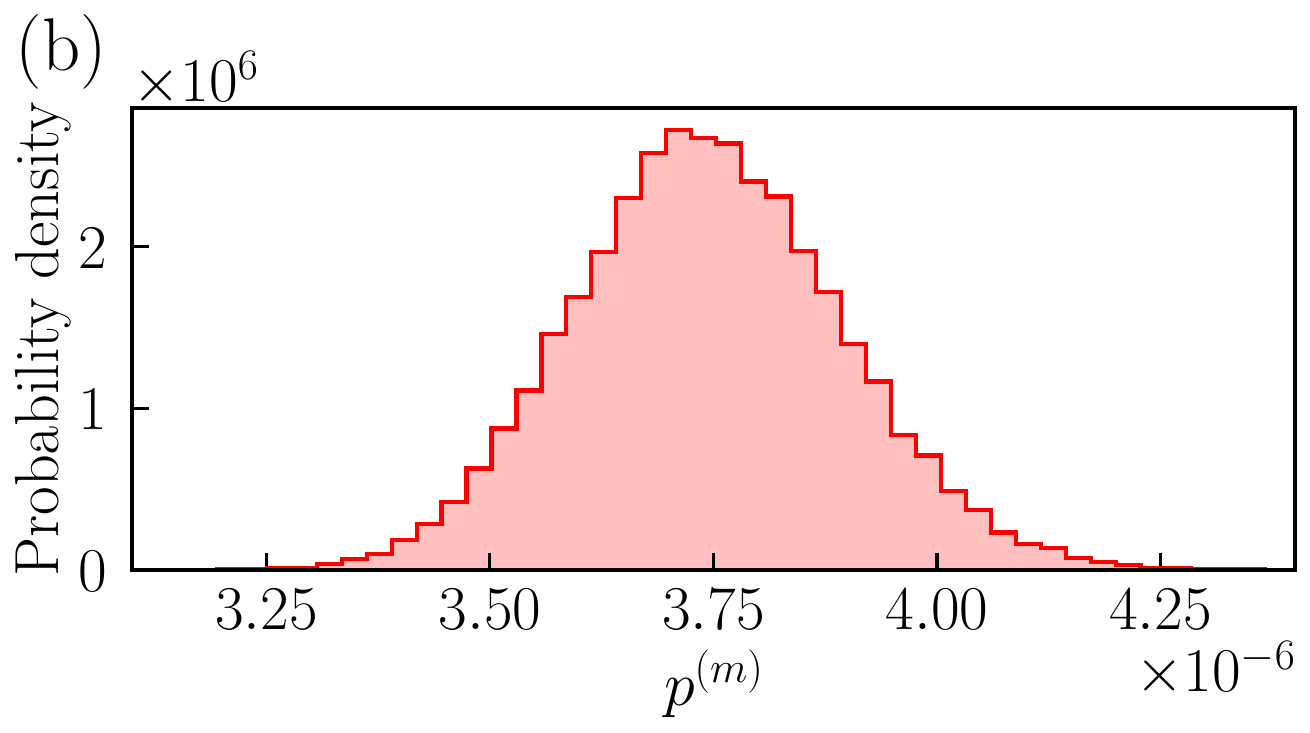}
\caption{Estimate of the statistical error using the blocking analysis
and the bootstrap method.
(a) At time $tJ=1$ for $N_{\mathrm{s}}=40$.
(b) At time $tJ=2N_{\mathrm{s}}$ for $N_{\mathrm{s}}=40$.
We choose $N_{\mathrm{total}}=2^{20}$, $N_{\mathrm{block}}=2^{10}$,
and $N_{\mathrm{boot}}=2^{12}$.
In both cases, the resampled data exhibit a normal distribution,
which allows us to estimate the statistical error safely.
}
\label{fig:3}
\end{figure}

Before estimating the expectation value and
the statistical error of
$\mathrm{perm}A$ in Eq.~\eqref{eq:rnd_smp_prod_rar},
we examine the distribution of the value $p^{(m)}$.
For simplicity, we will focus on the one-dimensional case for the moment.
We specifically consider the system size $N_{\mathrm{s}}=40$
and investigate the distribution of $2^{20}$ samples
at a short time ($tJ=1$) and at a long time ($tJ=2N_{\mathrm{s}}$).

Let us first look into the imaginary part of each sample
required for calculating $\mathrm{Im}\, \mathrm{perm}A$.
In the present study, we expect $\mathrm{Im}\, \mathrm{perm}A=0$
because $\mathrm{perm}A = \exp(-S_2)$ ($S_2\in \mathbb{R}$) should be real.
As shown in Fig.~\ref{fig:1},
we examine
the distribution of the positive and negative $\mathrm{Im}\, p^{(m)}$
at a short time $tJ=1$ and at a long time $tJ=2N_{\mathrm{s}}$.
In each time,
the positive and negative components
exhibit nearly the same shape
and cancel each other out, suggesting that
$\mathrm{Im}\, \mathrm{perm}A=0$ as expected
when the number of samples is sufficiently large.
Indeed, we numerically confirmed that the expectation value
of $\mathrm{Im}\, \mathrm{perm}A$ is always zero
within the sufficiently small statistical error.

We then investigate the real part of $\mathrm{perm}A$.
As in the case of imaginary part,
we examine
the distribution of the positive and negative $\mathrm{Re}\, p^{(m)}$
at a short time $tJ=1$ and at a long time $tJ=2N_{\mathrm{s}}$.

In the short time case ($tJ=1$),
we expect $S_2=\mathcal{O}(1)$ because
the entanglement entropy does not grow significantly.
Therefore, the condition	
$\mathrm{perm}A=\mathcal{O}(1)$
likely holds.
As we expected,
the distribution of the positive $\mathrm{Re}\, p^{(m)}$
has much greater weight than the negative one
[see Fig.~\ref{fig:2}(a)].
The positive component has a peak at $\mathrm{Re}\, p^{(m)} \approx + e^{-1}$,
whereas the negative component has a peak at $\mathrm{Re}\, p^{(m)} \approx - e^{-3}$.
Since the distribution is not a normal distribution,
we need careful analysis to estimate the statistical error,
as we will show later in this section.

However,
in the long time case ($tJ=2N_{\mathrm{s}}$),
we expect $S_2=\mathcal{O}(N_{\mathrm{s}})$
since the time-evolved state converges to
a highly entangled steady state.
Therefore, the condition
$\mathrm{perm}A=\mathcal{O}[\exp(-\mathrm{const.}\times N_{\mathrm{s}})]$
likely holds,
and the sampling must be much harder than the short time case.
Indeed, as shown in Fig.~\ref{fig:2}(b),
the positive and negative distributions exhibit a similar shape,
indicating that the expectation value is extremely small.
At the same time,
the area of the positive distribution is slightly larger
than that of the negative one, suggesting that
$\mathrm{Re}\, \mathrm{perm}A>0$.
As in the case of a short time,
the distribution of $\mathrm{Re}\, p^{(m)}$ is not a normal distribution,
which can be confirmed by the presence of
two peaks at $\mathrm{Re}\, p^{(m)} \approx \pm e^{-16}$.
Therefore, also for the long time case,
careful analysis is required to estimate the statistical error.

To estimate the statistical error, we combine the blocking analysis   
and the bootstrap method.
In the blocking analysis, we divide the $N_{\mathrm{total}}$ samples
in to the $N_{\mathrm{block}}$ blocks
containing $N_{\mathrm{blocksize}} = N_{\mathrm{total}}/N_{\mathrm{block}}$ samples.
For each block $j$ ($=1$, $2$, $\dots$, $N_{\mathrm{block}}$),
we calculate the average $p^{(j,N_{\mathrm{blocksize}})}$ of
$N_{\mathrm{blocksize}}$ samples.
This procedure results in
\begin{align}
 \frac{1}{N_{\mathrm{total}}}
 \sum_{m=1}^{N_{\mathrm{total}}}
 p^{(m)}
 &=
 \frac{1}{N_{\mathrm{block}}}
 \sum_{j=1}^{N_{\mathrm{block}}}   
 p^{(j,N_{\mathrm{blocksize}})},
\\
 p^{(j,N_{\mathrm{blocksize}})}
 &=
 \frac{1}{N_{\mathrm{blocksize}}}
 \sum_{k=(j-1)N_{\mathrm{blocksize}}+1}^{jN_{\mathrm{blocksize}}}
 p^{(k)}.   
\end{align}
We then prepare resampled data by the bootstrap method.
To this end,
we randomly choose $N_{\mathrm{block}}$ samples
$q^{(j)}$ ($j=1$, $2$, $\dots$, $N_{\mathrm{block}}$)
from the original $N_{\mathrm{block}}$ samples
$p^{(j,N_{\mathrm{blocksize}})}$ ($j=1$, $2$, $\dots$, $N_{\mathrm{block}}$).
Here, we do not avoid picking the same samples multiple times.
We repeat this process $N_{\mathrm{boot}}$ times
and generate samples
$\bar{q}^{(k)}$ ($k=1$, $2$, $\dots$, $N_{\mathrm{boot}}$)
by calculating
\begin{align}
 \bar{q}^{(k)} =
 \frac{1}{N_{\mathrm{block}}}
 \sum_{j=1}^{N_{\mathrm{block}}}
 q^{(j)}
\end{align}
for each $k$.
The number $N_{\mathrm{boot}}$ is chosen to be sufficiently large
so that the resampled data follows a normal distribution.
We estimate the average and the standard error of the samples $\bar{q}^{(k)}$,
which gives $\mathrm{perm} A$ and its statistical error
$\sigma_{\mathrm{perm} A}$.
Then, the statistical error of the R\'{e}nyi entanglement entropy is evaluated by
$\sigma_{S_2}
= | - \ln (\mathrm{perm} A + \sigma_{\mathrm{perm} A})
- [- \ln (\mathrm{perm} A) ] |
\approx |\sigma_{\mathrm{perm} A}/ \mathrm{perm} A|$
for $|\sigma_{\mathrm{perm} A}|\ll 1$.

In general, we do not need the blocking analysis; 
however, the computational cost of the bootstrap method will be extremely high
when we directly use the exponentially large number of $N_{\mathrm{total}}$ samples.
By taking a small constant $N_{\mathrm{block}}$,
we can reduce the computational cost of the bootstrap method.
Hereafter,
we typically choose $N_{\mathrm{block}}=2^{10}$ and $N_{\mathrm{boot}}=2^{12}$
and consider exponentially large $N_{\mathrm{total}}
\approx \exp(\mathrm{const.}\times N_{\mathrm{s}})$.
Note that the period of the pseudorandom number generator
should be sufficiently longer than the number of samples.
These parameters allow us to safely obtain a normal distribution
of the resampled data (for example, see Fig.~\ref{fig:3}).

\subsection{Size dependence of the statistical error}
\label{subsec:size_dep_error}

\begin{figure}[!t]
\centering
\includegraphics[width=\columnwidth]{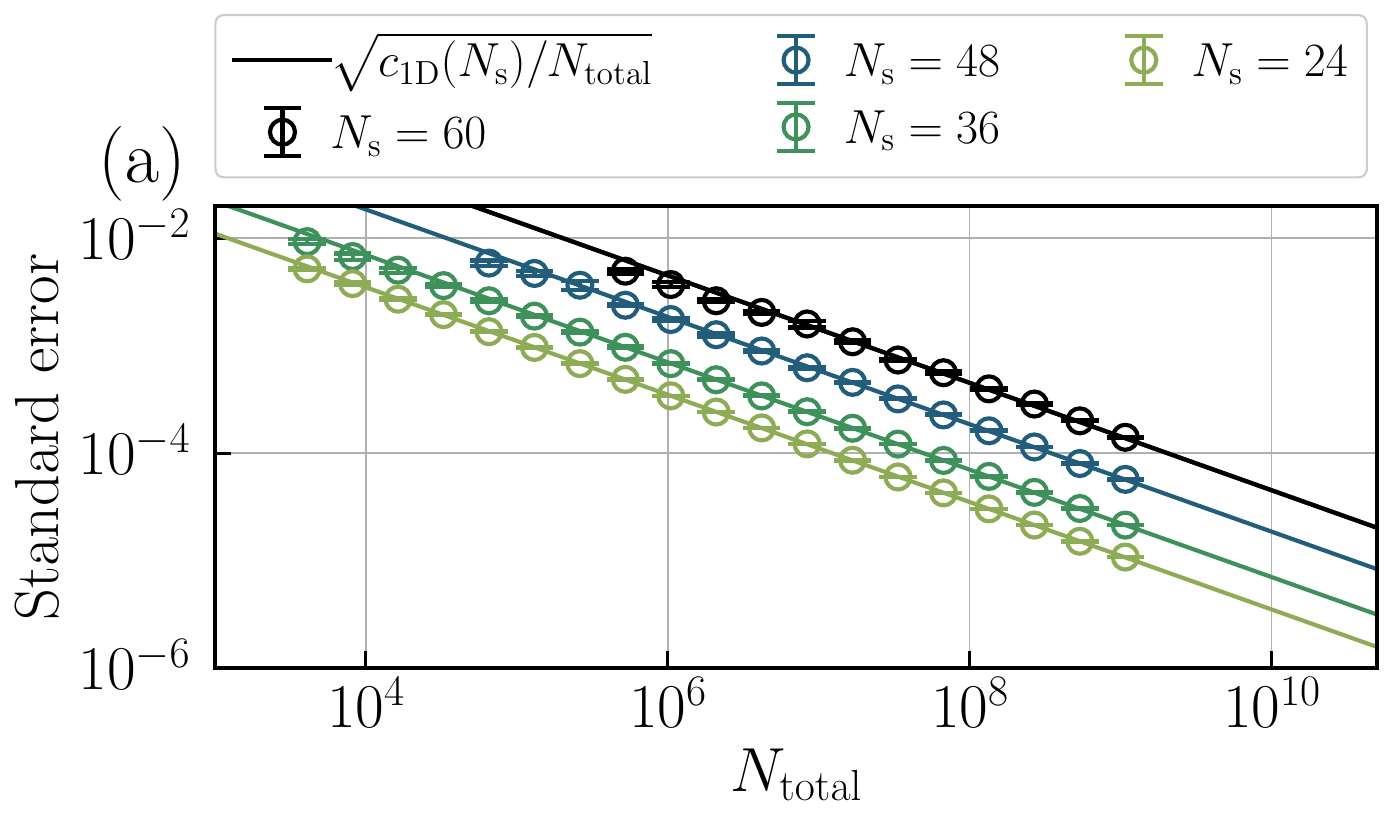}
\\
\includegraphics[width=\columnwidth]{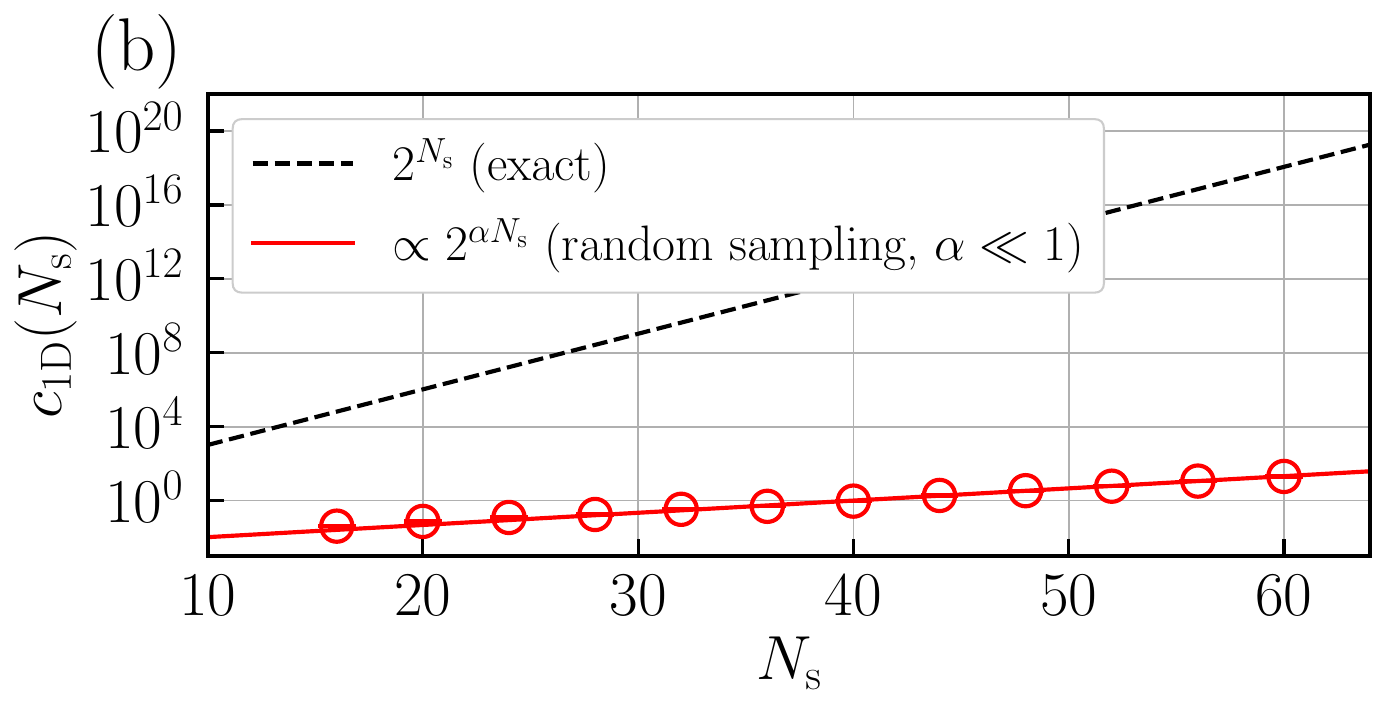}
\caption{(a) Size dependence of the standard error of R\'{e}nyi entanglement entropy density
$\sigma_{S_2/N_{\mathrm{s}}}$ at time $tJ=2N_{\mathrm{s}}$
as a function of the number of total samples $N_{\mathrm{total}}$ in 1D.
The statistical error is estimated by the blocking analysis
and the bootstrap method with $N_{\mathrm{block}}=2^{10}$
and $N_{\mathrm{boot}}=2^{12}$.
The error bar of $\sigma_{S_2/N_{\mathrm{s}}}$ is estimated for $32$ independent simulations.
The statistical error should satisfy
$\sigma_{S_2/N_{\mathrm{s}}} = \sqrt{c_{\mathrm{1D}}(N_{\mathrm{s}})/N_{\mathrm{total}}}$
with $c_{\mathrm{1D}}(N_{\mathrm{s}})$ being a size-dependent constant.
(b) Constant $c_{\mathrm{1D}}(N_{\mathrm{s}})$ as a function of size $N_{\mathrm{s}}$.
The value $c_{\mathrm{1D}}(N_{\mathrm{s}})$ represents the number of samples required
to achieve a given statistical error $\sigma_{S_2/N_{\mathrm{s}}}$.
We find that it satisfies
$c_{\mathrm{1D}}(N_{\mathrm{s}}) \approx 2^{0.2 N_{\mathrm{s}} - 9}$
by fitting data for $N_{\mathrm{s}}\ge 40$.
It is much smaller than the number of terms
($2^{N_{\mathrm{s}}}$)
in the summation in Eq.~\eqref{eq:bbfg},
suggesting that the computational cost is moderate
although it is exponential in $N_{\mathrm{s}}$.
}
\label{fig:4}
\end{figure}

\begin{figure}[!t]
\centering
\includegraphics[width=\columnwidth]{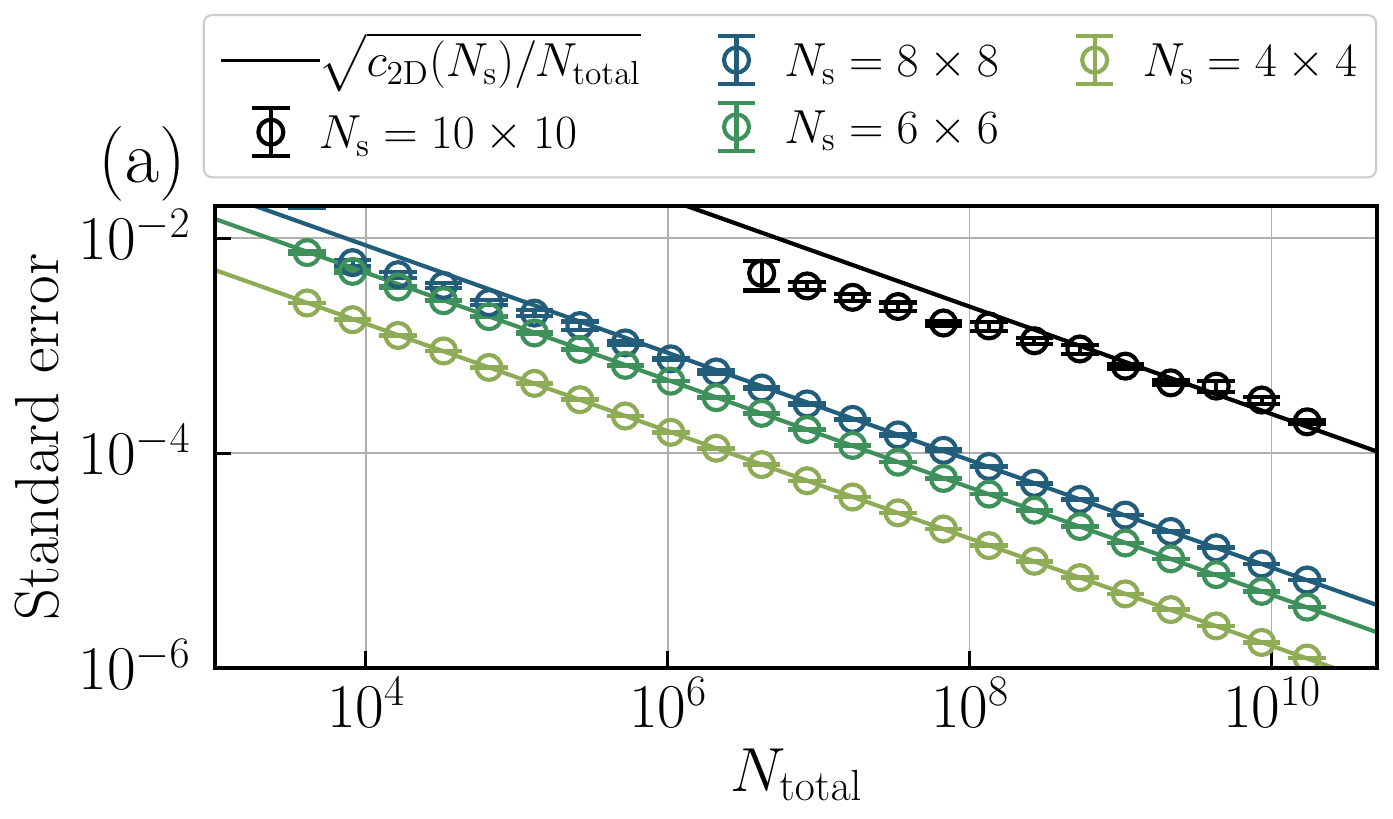}
\\
\includegraphics[width=\columnwidth]{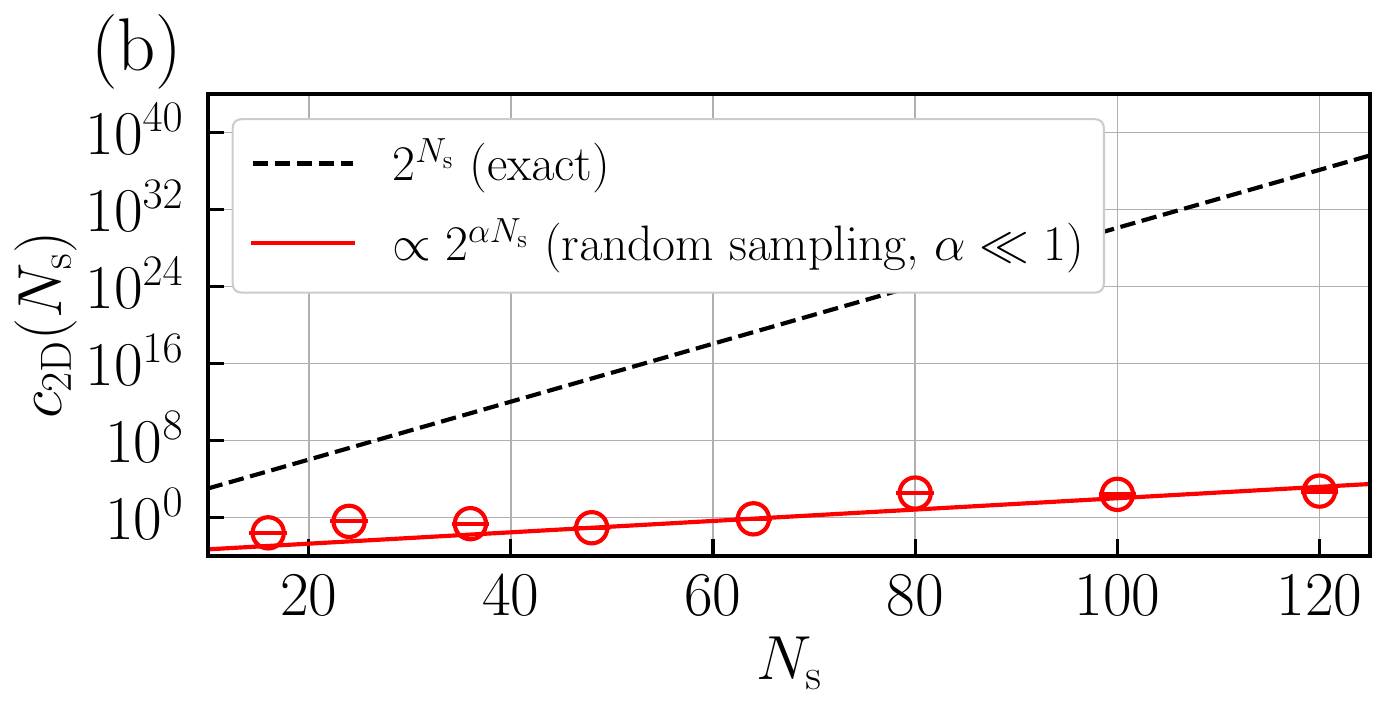}
\caption{(a) Size dependence of the standard error of R\'{e}nyi entanglement entropy density
$\sigma_{S_2/N_{\mathrm{s}}}$ at time $tJ=2L_x$
as a function of the number of total samples $N_{\mathrm{total}}$ in 2D.
We consider the lattice sites $N_{\mathrm{s}}=L_x\times L_y$
up to $N_{\mathrm{s}}=12\times 10$.
The statistical error is estimated by the blocking analysis
and the bootstrap method with $N_{\mathrm{block}}=2^{10}$
and $N_{\mathrm{boot}}=2^{12}$.
The error bar of $\sigma_{S_2/N_{\mathrm{s}}}$ is estimated for $32$ independent simulations.
The statistical error should satisfy
$\sigma_{S_2/N_{\mathrm{s}}} = \sqrt{c_{\mathrm{2D}}(N_{\mathrm{s}})/N_{\mathrm{total}}}$
with $c_{\mathrm{2D}}(N_{\mathrm{s}})$ being a size-dependent constant.
(b) Constant $c_{\mathrm{2D}}(N_{\mathrm{s}})$ as a function of size $N_{\mathrm{s}}$.
We find that it satisfies
$c_{\mathrm{2D}}(N_{\mathrm{s}}) \approx 2^{0.2 N_{\mathrm{s}} - 13}$
by fitting data for $N_{\mathrm{s}}>40$.
As in the case of 1D,
the prefactor ($\approx 0.2$) of $N_{\mathrm{s}}$ in 2D is much smaller than unity,
suggesting that the computational cost is moderate
although it is exponential in $N_{\mathrm{s}}$.
}
\label{fig:5}
\end{figure}

To estimate the ideal number of samples that we need for each system size,
we examine the size dependence of the number of samples
under the fixed statistical error.
Here, we focus on the one-dimensional system again.
For system sizes $N_{\mathrm{s}}=16$, $20$, $\dots$, $60$,
we increase the number of samples $N_{\mathrm{total}}$ up to $2^{30}$
and calculate the standard error of the R\'{e}nyi entanglement entropy density,
$\sigma_{S_2/N_{\mathrm{s}}}$.
The standard error for each system size $N_{\mathrm{s}}$ decreases as
\begin{align}
 \sigma_{S_2/N_{\mathrm{s}}}
 = \sqrt{\frac{c_{\mathrm{1D}}(N_{\mathrm{s}})}{N_{\mathrm{total}}}},
\end{align}
with increasing $N_{\mathrm{total}}$,
where $c_{\mathrm{1D}}(N_{\mathrm{s}})$ is a constant that depends on
$N_{\mathrm{s}}$ [see Fig.~\ref{fig:4}(a)].
The value $c_{\mathrm{1D}}(N_{\mathrm{s}})$ increases exponentially large
with increasing system size $N_{\mathrm{s}}$ in general.
By fitting numerical data points, we find
\begin{align}
 c_{\mathrm{1D}}(N_{\mathrm{s}})
 &= 2^{\alpha_{\mathrm{1D}} N_{\mathrm{s}} - \beta_{\mathrm{1D}}},
\\
 \alpha_{\mathrm{1D}} &= 0.219(6),
\\
 \beta_{\mathrm{1D}} &= 8.8(3),
\end{align}
as shown in Fig.~\ref{fig:4}(b).
This result suggests that the number of samples should be
\begin{align}
 N_{\mathrm{total}} = \frac{c_{\mathrm{1D}}(N_{\mathrm{s}})}{(\sigma_{S_2/N_{\mathrm{s}}})^2}
 \approx
 \frac{2^{0.2\times N_{\mathrm{s}} - 9}}{(\sigma_{S_2/N_{\mathrm{s}}})^2}
\end{align}
to keep the statistical error $\sigma_{S_2/N_{\mathrm{s}}}$ constant.
When we wish to suppress the statistical error,
e.g., $\sigma_{S_2/N_{\mathrm{s}}}=2^{-10}$,
the number of samples should be larger than
$N_{\mathrm{total}} = 2^{0.2N_{\mathrm{s}}+11}$.

The computational cost is 
proportional to the number of samples
and is
$\mathcal{O}(2^{\alpha_{\mathrm{1D}} N_{\mathrm{s}}})$
with $\alpha_{\mathrm{1D}}\approx 0.2\ll 1$
in the one-dimensional case.
Consequently, the random sampling method is much more efficient
than the conventional algorithms in Eq.~\eqref{eq:bbfg},
requiring the summation of $2^{N_{\mathrm{s}}}$
terms.

The similar small constant prefactor
$\alpha_{\mathrm{2D}}\approx 0.2$
is also found in the two-dimensional case
by analyzing systems up to $120$ sites.
As shown in Fig.~\ref{fig:5}(a),
we extract the size dependence of the coefficient
$c_{\mathrm{2D}} (N_{\mathrm{s}})$
in the fitting function
\begin{align}
 \sigma_{S_2/N_{\mathrm{s}}}
 = \sqrt{\frac{c_{\mathrm{2D}}(N_{\mathrm{s}})}{N_{\mathrm{total}}}}.
\end{align}
We find that the value $c_{\mathrm{2D}}(N_{\mathrm{s}})$
satisfies
\begin{align}
 c_{\mathrm{2D}}(N_{\mathrm{s}})
 &= 2^{\alpha_{\mathrm{2D}} N_{\mathrm{s}} - \beta_{\mathrm{2D}}},
\\
 \alpha_{\mathrm{2D}} &= 0.20(8),
\\
 \beta_{\mathrm{2D}} &= 13(4),
\end{align}
as shown in Fig.~\ref{fig:5}(b).
Therefore, the computational cost is also
$\mathcal{O}(2^{\alpha_{\mathrm{2D}} N_{\mathrm{s}}})$
with $\alpha_{\mathrm{2D}}\approx 0.2\ll 1$ in the
two-dimensional case.
In practice,
as for the system size $N_{\mathrm{s}}=10\times 10$
at the time point $tJ=20$,
it takes less than a day to calculate the R\'{e}nyi entanglement entropy
using a single core central processing unit.

\subsection{Entanglement entropy dynamics}
\label{subsec:ee_dynamics}

\begin{figure}[!t]
\centering
\includegraphics[width=\columnwidth]{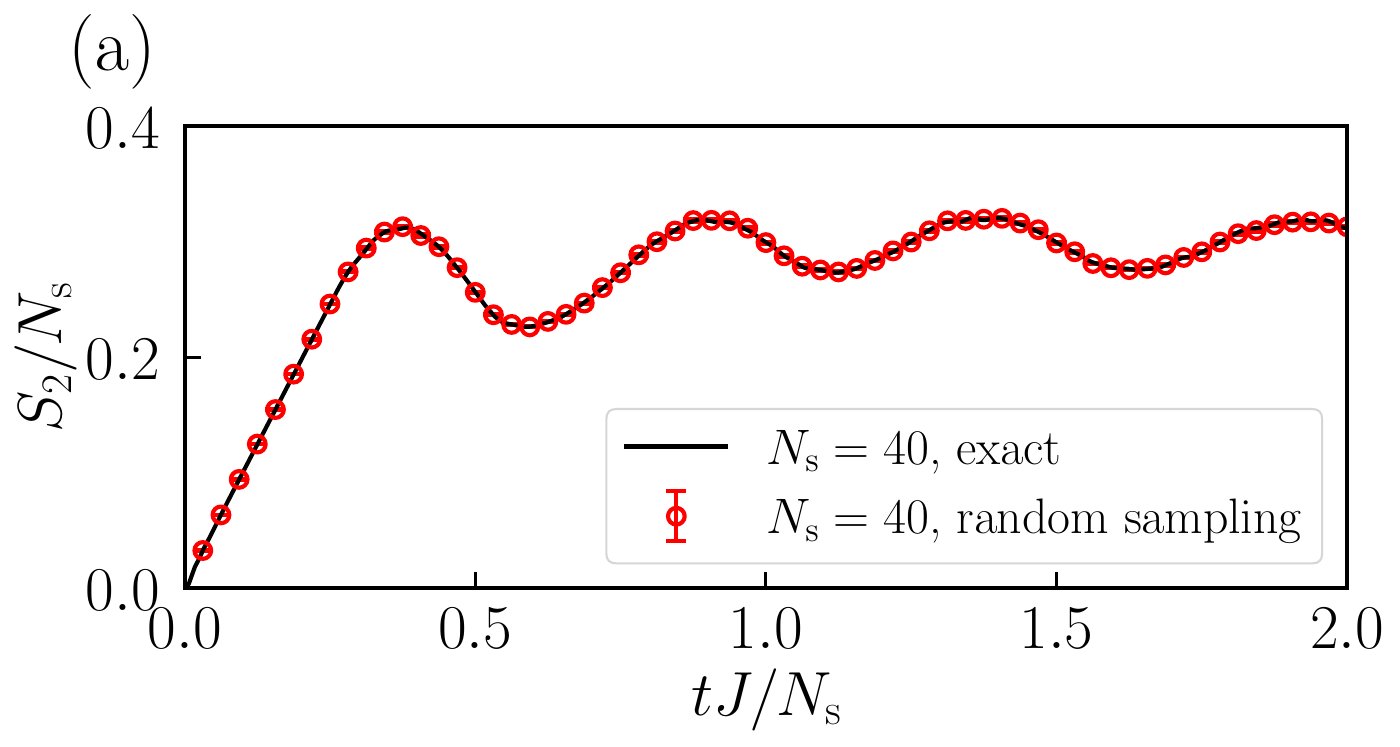}
\\
\includegraphics[width=\columnwidth]{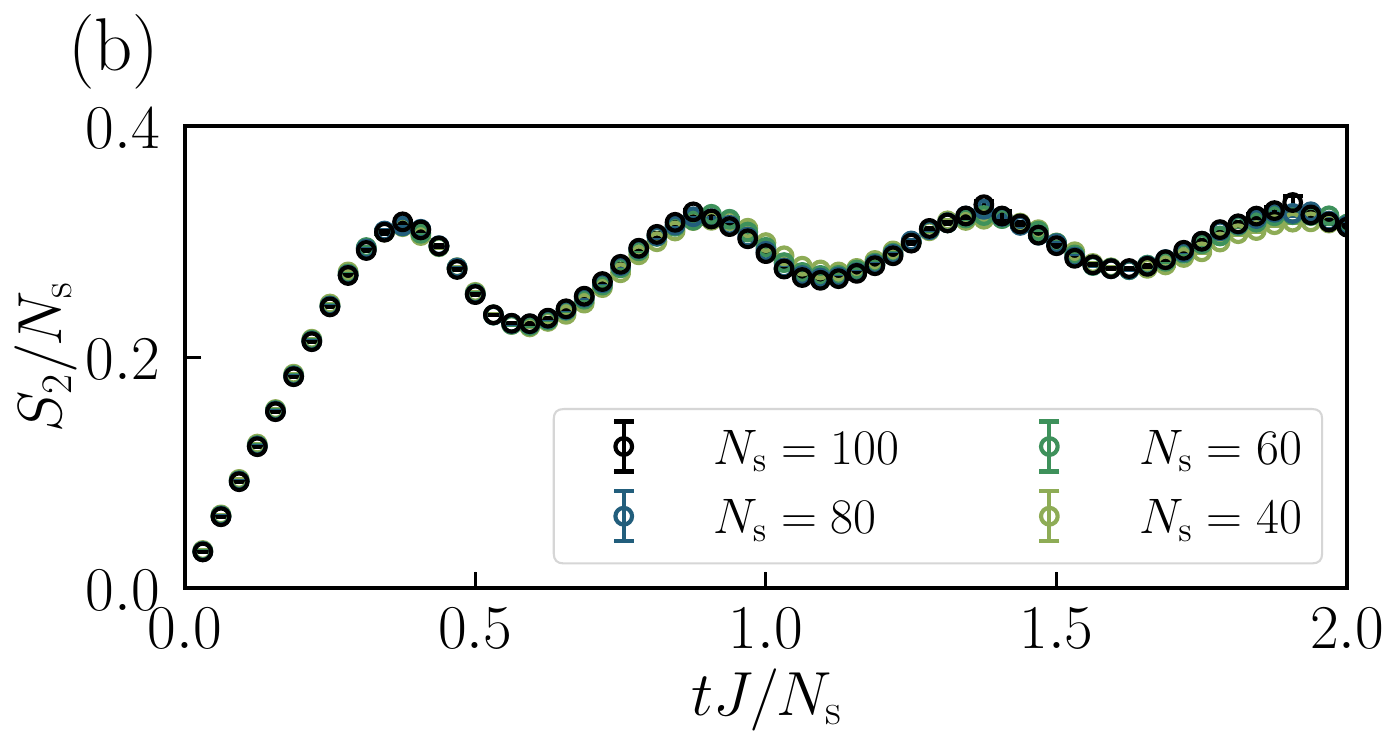}
\caption{(a) Comparison with the exact result
of the time evolution of the R\'{e}nyi entanglement entropy density
in 1D for $N_{\mathrm{s}}=40$,
which was the largest size obtained by the brute-force computation
of the matrix permanent.
The results are in good agreement.
(b) Time evolution of the R\'{e}nyi entanglement entropy density
for much larger systems.}
\label{fig:6}
\end{figure}

\begin{figure}[!t]
\centering
\includegraphics[width=\columnwidth]{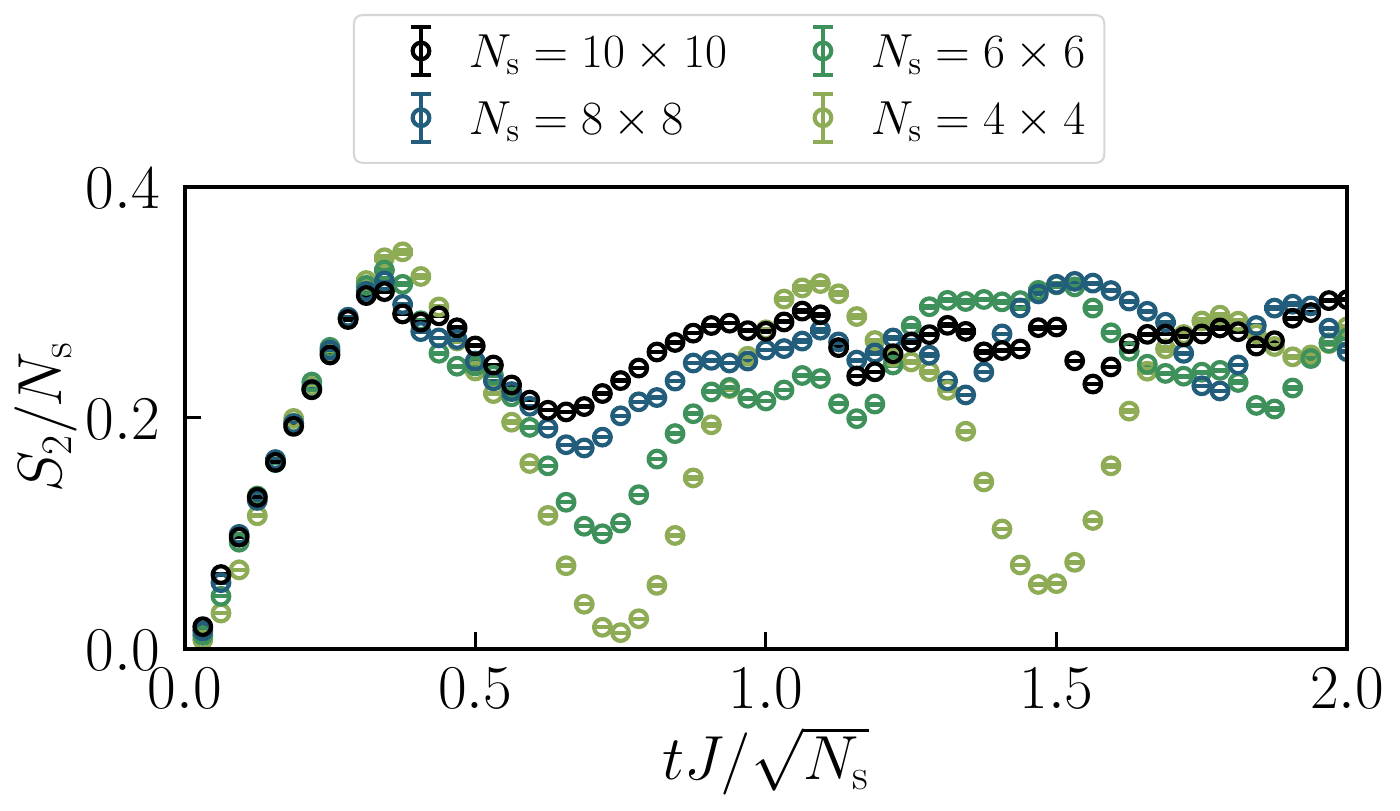}
\caption{Time evolution of the R\'{e}nyi entanglement entropy density in 2D.
We consider the lattice sites up to $N_{\mathrm{s}}=10\times 10$
and calculate the R\'{e}nyi entanglement entropy density
when the system is divided into identical two parts.}
\label{fig:7}
\end{figure}

By taking advantage of the random sampling method,
we calculate the dynamics of R\'{e}nyi entanglement entropy density
after a sudden quench.
Hereafter, we choose the number of samples
$N_{\mathrm{total}} = 2^{0.2N_{\mathrm{s}}+12}$
to keep the statistical error sufficiently small.

Let us first compare our present result with the exact one
calculated with the largest size $N_{\mathrm{s}}=40$
in our previous study~\cite{kagamihara2023}
in the case of 1D.
As shown in Fig.~\ref{fig:6}(a),
the random sampling method provides the exact R\'{e}nyi entanglement entropy density
within the statistical error bar.

We then study the larger systems up to $N_{\mathrm{s}}=100$.
As shown in Fig.~\ref{fig:6}(b),
the error bar is sufficiently small for all sizes that we study.
The R\'{e}nyi entanglement entropy densities for $N_{\mathrm{s}}\ge 40$ nearly overlap, exhibiting the volume law scaling.
Thus, the system size $N_{\mathrm{s}}=40$,
corresponding to the largest size in our previous study,
is large enough to capture the nature of the entanglement entropy density dynamics
in the thermodynamic limit.

Next, we investigate the R\'{e}nyi entanglement entropy density dynamics
in a two-dimensional square lattices.
As shown in Fig.~\ref{fig:7},
the R\'{e}nyi entanglement entropy density grows linearly in time
for a short time up to $tJ \approx 0.3 \sqrt{N_{\mathrm{s}}}$
for $N_{\mathrm{s}} = L_x \times L_y$ with $L_x=L_y$.
The behavior is consistent with the prediction
from the previous studies on the entanglement entropy density dynamic
in integrable systems
with the Gaussian initial states~\cite{alba2017,alba2018},
although our initial state is not the Gaussian state.
The system-size dependence of the entanglement entropy density dynamics
is rather small in this time regime.
When the time is longer than $tJ \approx 0.3 \sqrt{N_{\mathrm{s}}}$,
the entanglement entropy density shows a larger size dependence.
It is difficult to extract the physically meaningful interpretation
of the entanglement entropy density dynamics in the thermodynamic limit.
However,
as the system size increases,
the fluctuation of the entanglement entropy density becomes smaller.
The entanglement entropy density
appears to converge to a certain value,
exhibiting volume-law behavior of the entanglement entropy consistent with
the previous studies~\cite{alba2017,alba2018}.
Within the system sizes that we study,
the entanglement entropy density approximately approaches
the value close to $\approx 0.3$ in both 1D and 2D.

\section{Conclusions and outlook}
\label{sec:conclusions}

In conclusion,
we studied the dynamics of the R\'{e}nyi entanglement entropy
of insulating initial states in free boson systems.
Owing to the non-Gaussian nature of the initial states,
the calculation of the entanglement entropy required the evaluation
of the matrix permanent, which has the exponential cost.
We developed a random sampling method for evaluating the matrix permanent
and found that the computational cost was reduced to
$\mathcal{O}(2^{\alpha N_{\mathrm{s}}})$
with a small constant $\alpha \approx 0.2 \ll 1$
in one-dimensional and two-dimensional
$N_{\mathrm{s}}$-site systems at half filling.
This reduction enabled us to study the entanglement entropy dynamics
for more than $100$ sites in free boson systems.

Our results can be tested in experiments involving ultracold atoms in
optical lattices and trapped ions. The dependence of entanglement
entropy dynamics on system size is weak for
one-dimensional systems with more than $40$ sites;
however, the dynamics have not converged even with $100$ sites in
2D. Although performing sufficiently large-scale quantum simulations
with current techniques remains challenging, it would be valuable to
qualitatively verify the dependence of entanglement entropy dynamics on
spatial dimensions. Our numerical data will assist in comparing these
experimental results.

In the present study,
we applied the simple random sampling method
to the calculation of the R\'{e}nyi entanglement entropy.
One may consider more sophisticated sampling methods,
such as the rejection sampling method~\cite{huber2008,kuck2019,harviainen2021}
and the importance sampling method~\cite{jerrum2004,bezakova2008,kou2009},
to reduce the variance of the estimator
for the matrix permanent.
The upper and lower bounds of the entanglement entropy,
i.e., the lower and upper bounds of the matrix permanent,
would be utilized during such sophisticated sampling.
As for the upper bound of the entanglement entropy,
the second R\'{e}nyi entanglement entropy
is bounded above by the von Neumann entanglement entropy,
and the von Neumann entanglement entropy is
bounded above by the von Neumann entanglement entropy
of a certain Gaussian state having the same two-point correlation
functions as the original state~\cite{bianchi2018,hackl2018}.
The entanglement entropy of the Gaussian state
can often be calculated efficiently.
As for the lower bound of the entanglement entropy,
by utilizing the following inequalities~\cite{berkowitz2018}
for the matrix $A$
in Eq.~\eqref{eq:ee_a}
that always fulfills $||A||_2=1$,
with $|| \cdot ||_2$ being the operator $2$-norm~\cite{kagamihara2023}:
\begin{align}   
 \mathrm{perm} A
 &\le
 \mathbb{E}\left[\prod_{i=1}^{n} \left| r^{*}_i
 \left( \sum_{j=1}^{n} a_{ij} r_j \right) \right| \right]
 =
 \mathbb{E}\left[\prod_{i=1}^{n}
 \left| \sum_{j=1}^{n} a_{ij} r_j \right| \right]
\\
 &\le  
 \mathbb{E}\left[
 \left(
 \frac{1}{n} \sum_{i=1}^{n}
 \left| \sum_{j=1}^{n} a_{ij} r_j \right|
 \right)^n
 \right]
\\
 &\le
 \mathbb{E}\left[
 \left(
 \frac{1}{\sqrt{n}}
 \sqrt{
 \sum_{i=1}^{n}
 \left| \sum_{j=1}^{n} a_{ij} r_j \right|^2
 }
 \right)^n
 \right]
\\
 &\le
 \mathbb{E}\left[
 ||A||_2
 \right]
 = 1,
\end{align}
one may consider the entanglement-entropy-like quantities,
\begin{align}
\label{eq:ee_lower_bound_s2_1}
 S_2'
 &=
 -\ln
 \mathbb{E}\left[\prod_{i=1}^{n}
 \left| \sum_{j=1}^{n} a_{ij} r_j \right| \right],
\\
\label{eq:ee_lower_bound_s2_2}
 S_2''
 &=
 -\ln
 \mathbb{E}\left[
 \left(
 \frac{1}{n} \sum_{i=1}^{n}
 \left| \sum_{j=1}^{n} a_{ij} r_j \right|
 \right)^n
 \right],
\\
\label{eq:ee_lower_bound_s2_3}
 S_2'''
 &=
 -\ln
 \mathbb{E}\left[
 \left(
 \frac{1}{\sqrt{n}}
 \sqrt{
 \sum_{i=1}^{n}
 \left| \sum_{j=1}^{n} a_{ij} r_j \right|^2
 }
 \right)^n
 \right],
\end{align}
satisfying
\begin{align}
 S_2 \ge S_2' \ge S_2'' \ge S_2''' \ge 0.
\end{align}
The quantities $S_2'$, $S_2''$, and $S_2'''$
can be calculated more efficiently than $S_2$
using the simple random sampling method
or the importance sampling method
because the quantities inside the expectation operator $\mathbb{E}$
are always nonnegative.
When we wish to apply the rejection sampling method,
for example, we may utilize the relation
between the quantities
$p(\bm{r})
= \prod_{i=1}^{n} {r_i}^{*}
\left( \sum_{j=1}^{n} a_{ij} r_j \right)$
in Eq.~\eqref{eq:rnd_smp_prod_rar_p}
and
$q(\bm{r})
:= \prod_{i=1}^{n}
\left| \sum_{j=1}^{n} a_{ij} r_j \right|$
that appears in Eq.~\eqref{eq:ee_lower_bound_s2_1}.
Since $q(\bm{r})$ is always nonnegative
and the relation $p(\bm{r})\le q(\bm{r})$ holds
for any $\bm{r}$,
we can sample the random vector $\bm{r}$
from the distribution that generates $q(\bm{r})$
using the simple random sampling method
and then sample $\tilde{s}$
from the uniform distribution on the interval $[0,q(\bm{r})]$.
The sample $\bm{r}$ is accepted if $\tilde{s}\le p(\bm{r})$
and is rejected otherwise.
One can also combine the rejection and importance sampling
methods~\cite{liu1998}.
The negative-sign-problem-like difficulty would be
slightly alleviated when $p(\bm{r})$ is close to $q(\bm{r})$
for $\bm{r}$ that is likely to be sampled.

Although we specifically focused on the $010101\cdots$-type CDW initial
state, our approach can apply to other initial states
that can be represented by a simple product of local Fock states.
When using other initial states
where the number of particles at each site is either $0$ or $1$,
one has to appropriately modify the set
$\mathrm{G}_{\mathrm{CDW}}$
of charge rich sites in Eq.~\eqref{eq:cdw_init}.
When initial states have two or more particles at each site,
the situation is more complex, although it is possible to extend the
formalism using similar calculations.
The random sampling method is also applicable to the dynamics
of the entanglement entropy in
general noninteracting Hamiltonians
including long-range and random hopping terms.
Such Hamiltonians only modify their eigenenergies and
eigenstates defined in Eq.~\eqref{eq:ham_0}.

We expect that the computational cost of the random sampling method
does not significantly depend on the details of the initial states
for the parameter range that exhibits the volume-law scaling
of the entanglement entropy.
When the initial state contains $N_{\mathrm{b}}$ particles,
we need to evaluate the permanent of an $N\times N$ matrix
with $N=2N_{\mathrm{b}}$ to calculate the entanglement
entropy.
We speculate that
the factor $\alpha$ in the computational cost
$\mathcal{O}(2^{\alpha N})$ of the random sampling method would be
primarily determined by the size of the entanglement entropy
per particle.
This is because when the entanglement entropy per particle $s$
is small and close to zero, the sample
$\mathrm{Re}\, p^{(m)}[\approx \exp(-s N)]$
in Eq.~\eqref{eq:rnd_smp_prod_rar_p}
should be close to unity for most samples $m$,
indicating that the most of the samples are positive
[see Fig.~\ref{fig:2}(a) as an example in the case of $s\approx 0$].
Consequently, the sampling efficiency increases
and the factor $\alpha$ decreases,
irrespective of the choice of the initial state
as long as the entanglement entropy per particle is $s$.

As for the one-dimensional and two-dimensional systems
that we study,
the entanglement entropy per particle
is $s\approx 2\times 0.3$,
whereas the corresponding factor is
$\alpha\approx 0.2\ll 1$.
Our finding suggests that
the factor $\alpha$ would be smaller than unity
even when the entanglement entropy per particle is
$\mathcal{O}(1)$,
which is the case in physically relevant systems
exhibiting volume-law scaling of the entanglement entropy.
This is in contrast to the conventional algorithms
that always require the summation of $2^{N}$ terms,
corresponding to the case of $\alpha=1$
in the random sampling method.
It is intriguing to explore
how the computational cost of the random sampling method
depends on the entanglement entropy per particle
in various initial states
and in other noninteracting systems.

We specifically studied the dynamics of the R\'{e}nyi entanglement
entropy in free boson systems after a sudden quench.
One may also consider the problems in the boson sampling
devices~\cite{aaronson2011,aaronson2014}.
There are several proposals for reducing the computational cost
of the matrix permanent regarding
the boson sampling procedure~\cite{quesada2022,clifford2024}.
In practice, the feasible matrix size is up to
$\approx 50\times 50$ so far~\cite{neville2017,wu2018,lundow2022}.
It is an interesting problem to study whether the sampling method
also reduces the computational cost of the permanent of
the matrix representing the boson sampling task.

As for the dynamics in the presence of interactions,
namely, the dynamics in the Bose-Hubbard model,
the information propagation
and the particle transport
would behave differently~\cite{kuwahara2024}.
The former speed could be much faster than the latter speed.
Since the numerical investigation of the entanglement entropy dynamics
in strongly correlated systems is much more challenging,
studying the dynamics in noninteracting boson systems
using the random sampling method
would help understand the information propagation
in nonequilibrium quantum systems.

\begin{acknowledgments}
The authors acknowledge fruitful discussions with
Shimpei Goto,
Kota Sugiyama,
Yuki Takeuchi,
Shunji Tsuchiya,
Shion Yamashika,
and
Ryosuke Yoshii.
This work was supported by
JSPS KAKENHI (Grant No.\ JP24H00973),
MEXT KAKENHI, Grant-in-Aid for Transformative Research Area
(Grants No.\ JP22H05111 and No.\ JP22H05114),
JST FOREST (Grant No.\ JPMJFR202T),
and
MEXT Q-LEAP (Grant No.\ JPMXS0118069021).
The numerical computations were performed on computers at 
the Yukawa Institute Computer Facility,
Kyoto University
and on computers at
the Supercomputer Center, the Institute for Solid State Physics,
the University of Tokyo.
\end{acknowledgments}

\onecolumngrid

\end{document}